\documentclass[a4paper,fleqn]{cas-sc}

\usepackage[numbers]{natbib}
\usepackage{multirow}
\usepackage{makecell}
\usepackage{caption,booktabs}
\usepackage{tabularx}

\def\tsc#1{\csdef{#1}{\textsc{\lowercase{#1}}\xspace}}
\tsc{WGM}
\tsc{QE}

\graphicspath{{figures/}}

\usepackage[utf8]{inputenc}
\usepackage{graphicx} 
\usepackage{xcolor}

\usepackage{tabularx}
\usepackage{listings}
\usepackage{color}

\definecolor{mygreen}{rgb}{0,0.6,0}
\definecolor{mygray}{rgb}{0.5,0.5,0.5}
\definecolor{mymauve}{rgb}{0.58,0,0.82}
\definecolor{backcolour}{rgb}{0.95,0.95,0.92}

\begin{document}
\let\WriteBookmarks\relax
\def\floatpagepagefraction{1}
\def\textpagefraction{.001}

\shorttitle{A Framework to Model ML Engineering Processes}    

\shortauthors{Morales S., Clarisó R., Cabot J.}  

\title [mode = title]{A Framework to Model ML Engineering Processes}  

\author[1]{Sergio Morales}[orcid=0000-0002-5921-9440]
\cormark[1]
\ead{smoralesg@uoc.edu}

\author[1]{Robert Clarisó}[orcid=0000-0001-9639-0186]
\ead{rclariso@uoc.edu}

\author[2,3]{Jordi Cabot}[orcid=0000-0003-2418-2489]
\ead{jordi.cabot@list.lu}

\affiliation[1]{organization={Universitat Oberta de Catalunya},
            addressline={Rambla del Poblenou 156}, 
            postcode={08018},
            city={Barcelona},
            country={Spain}}

\affiliation[2]{organization={Luxembourg Institute of Science and Technology},
            addressline={5 Av. des Hauts-Forneaux},
            postcode={4362},
            city={Esch-sur-Alzette},
            country={Luxembourg}}
\affiliation[3]{organization={University of Luxembourg},
            addressline={2 Rue de l’Université},
            postcode={4365},
            city={Esch-sur-Alzette},
            country={Luxembourg}}

\begin{abstract}
The development of Machine Learning (ML) based systems is complex and requires multidisciplinary teams with diverse skill sets. This may lead to communication issues or misapplication of best practices. Process models can alleviate these challenges by standardizing task orchestration, providing a common language to facilitate communication, and nurturing a collaborative environment. Unfortunately, current process modeling languages are not suitable for describing the development of such systems. In this paper, we introduce a framework for modeling ML-based software development processes, built around a domain-specific language and derived from an analysis of scientific and gray literature. A supporting toolkit is also available.
\end{abstract}


\maketitle

\section{Introduction}
\label{sec:intro}
In the last years, Artificial Intelligence (AI) has rapidly become a key element of most software systems (sometimes referred to as \emph{smart} software systems) as business applications embed Machine Learning (ML) and other AI components as core aspects of their logic \citep{Anthes,Deng}. 

Therefore, modern software teams are evolving to become multidisciplinary and often include other profiles beyond developers and software engineers, \emph{e.g.}, data scientists, psychologists, and AI experts. The need for proper guidance when executing a project to develop AI-based software and seamlessly integrate such a diverse skill set is manifest in recent studies with practitioners \citep{Amershi,Bosch,Hill,Wan}. 

This guidance can be provided by process models. A \emph{process model} provides full visibility and traceability about the work decomposition within an organization, along with the responsibilities of their participants and the standards and knowledge it is based on. Process models are guidelines for configuration, execution and continuous improvement. Software process models are specific types of process models focused on the specification of the development process to follow to create a new software artifact. Nevertheless, current process modeling languages are not suited for defining the new types of activities, roles and best practices required in the development of smart software systems.

In this sense, this article proposes a new framework to provide appropriate guidance for modeling and enacting ML-based software engineering processes. This modeling framework is built around a \emph{domain-specific language} (DSL) that combines standard process modeling concepts with AI-specific process primitives. This DSL is based on the analysis of research and gray literature on AI engineering.

A DSL provides a shared language in a particular problem space that fosters communication and collaboration between all stakeholders. A preliminary version of the DSL was published in \citet{Morales} and has been extended in this work in several aspects. Firstly, we have enriched the DSL for contextualizing and aligning the AI aspects with the business purpose. Moreover, the DSL now also enables users to define the operationalizing and monitoring aspects of the development process. Furthermore, a modeling editor is provided to facilitate process editors to model processes in a seamless graphical way. A transformation of our language to BPMN~\citep{BPMN} is also provided to facilitate the integration of our work with other BPMN compliant tools. Finally, an HTML documentation generator is included in the toolkit.

We expect our framework to have several benefits in an organization. Among them, the simplification of the on-boarding of new team members and enabling them to seamlessly create their first consistent AI application. The DSL is also a foundational step towards automatic processing, \emph{e.g.}, as part of a process execution scenario. A process model built with our DSL may detect hidden, implicit or conflicting activities in already existing software processes. We consider that our approach will enable the standardization of practices by providing a structured modeling framework so that process experts can effectively describe, promulgate and execute their company's own AI engineering processes.

The remainder of the article is organized as follows. Section~\ref{sec:modelingframework} 
introduces the process modeling framework and its components. Section~\ref{sec:mlenginpractice} presents the needs that arose from the industry and the current available methods to tackle them and that feed the design of our DSL, which is afterward exposed in Section~\ref{sec:dsldesign}, where we define its elements and semantics. In Section~\ref{sec:toolkit} we present the developed modeling editor which is based on the proposed DSL, and we introduce additional features of the toolkit, whereas in Section~\ref{sec:application} we illustrate their application scenarios. Section~\ref{sec:relatedwork} explores frameworks 
proposed for addressing problems regarding the definition and execution of processes. Finally, Section~\ref{sec:conclusions} concludes and outlines the future work.

\section{Overview}
\label{sec:modelingframework}
\begin{figure}[t]
    \includegraphics[width=\textwidth]{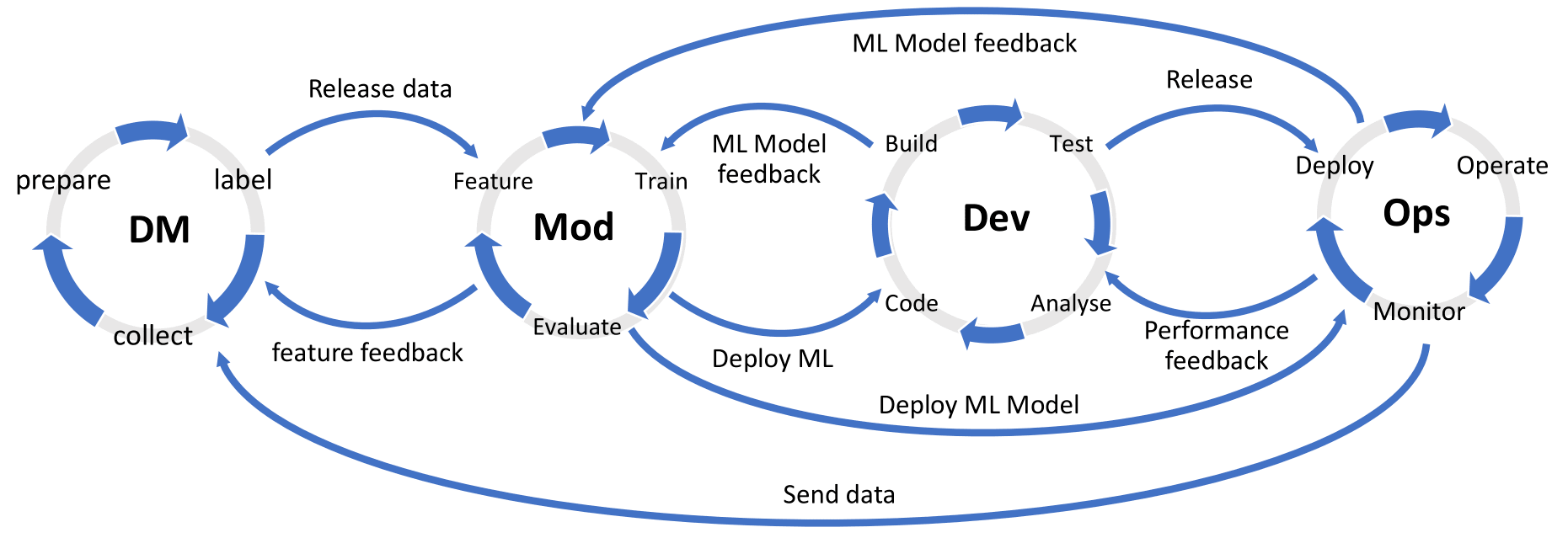}
    \caption{ML workflow and DevOps process integration, from \citet{Lwakatare, Tamburri}.}
    \label{fig:MLOpsLifecycle}
\end{figure}
Our modeling framework 
considers the needs of the industry in developing AI/ML software and puts forward a toolkit to seamlessly execute AI/ML projects end-to-end\footnote{Our DSL covers generic aspects of AI technologies, and ML in particular.}. A typical AI-based software modern development process is often illustrated as ML and DevOps workflows adjoined, as in Figure~\ref{fig:MLOpsLifecycle}.

Our aim is to guide practitioners in their duties of \emph{designing}, \emph{enacting} and \emph{documenting} their organization's internal processes. The main objective of this framework is to help in the definition and work breakdown structure of the ML specific activities usually performed to develop a successful AI-based software product within an organization. It also helps to capture their orchestration, bearing in mind that the end user's technical knowledge might be limited. The framework includes the identification of roles and their functions, along with the assignment of responsibilities towards the different activities of the new system. It provides the description of the usual entities, their attributes, associations, and constraints (graphic and textual), necessary to carry out the AI specific activities and model their results. Ultimately, the framework enables a process modeler to realize their work using a model-driven paradigm.

In short, the framework provides a systematic environment for holistically modeling AI software development processes, taking into account the semantics associated with the AI specific activities, the resources consulted and artifacts generated, and the roles that participate in the activities. The modeling is done independently of the underlying technological platform and the convenience of the different components and tools is described, especially that of the workflow activities, as it is the central axis of the framework.

\begin{figure*}
\centering
\includegraphics[width=\textwidth]{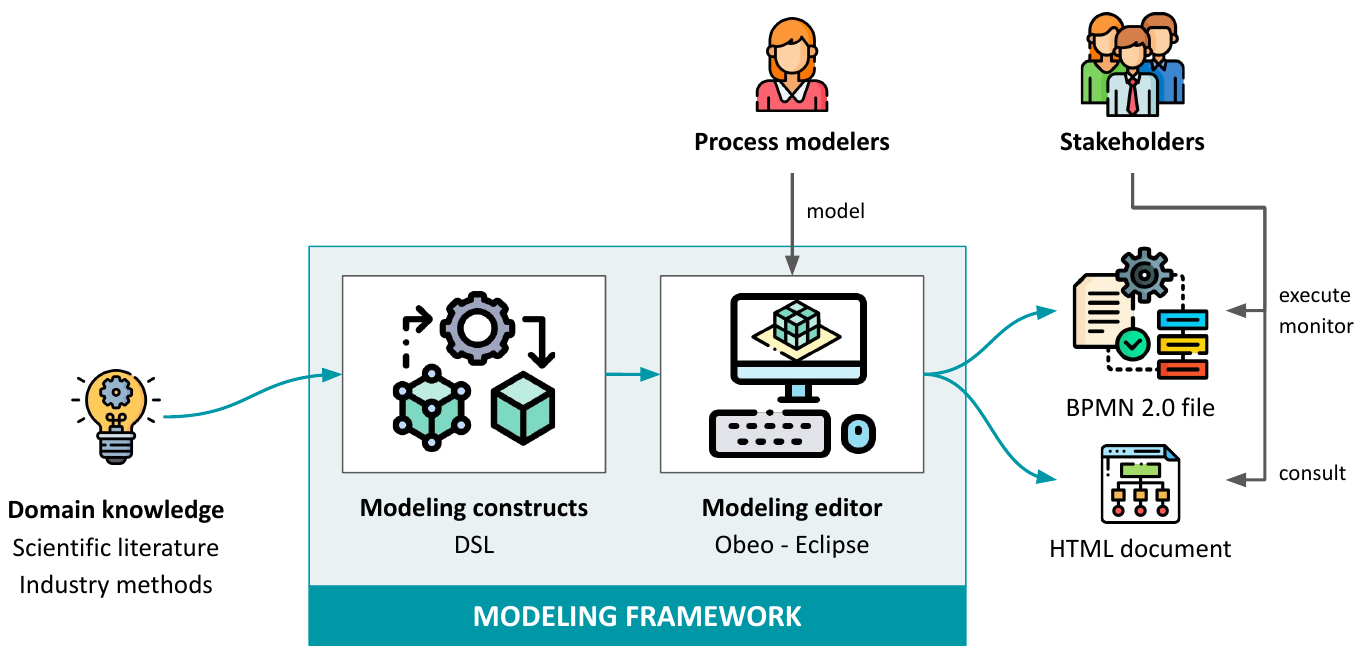}
\caption{The components of the framework and its toolkit.}
\label{fig:ModelingFrameworkToolkit}
\end{figure*}

As we see in Figure~\ref{fig:ModelingFrameworkToolkit}, the modeling framework contributes a \emph{domain-specific language} (DSL) that describes the syntax and semantics of the modeling assets {\textendash}activities, roles, resources and other artifacts{\textendash}, and the rules that govern them. The elements of the DSL {\textendash}thoroughly explored in Section~\ref{sec:dsldesign}{\textendash} could be assembled in order to build up an AI software development process specific to a given organization. A process modeler could design a process using the \emph{modeling editor}, which is based on the DSL and thus provides the constructs and semantics to rigorously perform this task. The implementation and its usage hints are presented in Section~\ref{sec:modelingeditor}. As a practical example of the applicability of the modeling editor, we have accompanied it with a \emph{BPMN converter} {\textendash}described in Section~\ref{sec:bpmnconverter}{\textendash} to export the modeled process into a BPMN 2.0 compliant file that could be imported into and executed by any standard compatible BPMN platform. We have also included a plug-in to generate HTML documentation based on the modeled process {\textendash}featured in Section~\ref{sec:htmlgenerator}{\textendash} so that any employee in the organization is able to browse the information contained.

Note that our framework does not prescribe any concrete AI engineering process model. Instead, it offers the required support so that each organization can easily define its own process.

\section{ML Engineering in Practice}
\label{sec:mlenginpractice}

\emph{Machine Learning Engineering} is commonly defined as the mixed use of principles, tools, and techniques from both Machine Learning and software engineering areas to design and build AI-based systems \citep{Burkov}. Particularly, agile practices are often applied to develop modern software, and one of the most used agile approach nowadays is DevOps. The DevOps process model tightens the collaboration between development and operation of software systems. \emph{MLOps} is defined as \emph{"an ML engineering culture and practice that aims at unifying ML system development (Dev) and ML system operation (Ops)"} \citep{GoogleCloud}. Other authors describe it as an extended version of DevOps complemented by ML-related roles and activities to deploy and operationalize Machine Learning models in production \citep{Alla,Lwakatare,Martinez-Fernandez1,Shankar}.

We have found a number of papers exposing the ML engineering process they followed in their individual scenario or illustrating a generic approach \citep{Akkiraju,Amershi,Ashmore,Bosch,Bosch2018,Haakman,Hesenius,Khomh,Kreuzberger,Lwakatare,Moreb,Nascimento,Ranawana,Shankar,Vaidhyanathan}. We have scrutinized them to make sure all the aspects they mention are part of both the references that discuss the problematics and challenges of such ML engineering efforts \citep{Alvarez-Rodriguez,Amershi,Bosch,Colomo-Palacios,Liu,Lwakatare2019,Nascimento,Sculley,Serban,Wan,Nahar} and the literature reviews that summarize them \citep{Giray,Kumeno,Martinez-Fernandez,Nascimento2}.

Among those first aforementioned, we have selected the most descriptive research publications, and complemented them with influential contributions from the industry, as the foundations for building up our language. We have not considered short papers, nor scientific publications that we found to either (1) propose a subset of previous findings, (2) put forward domain-specific solutions, or (3) be extremely brief in their definitions. We wanted to make sure our DSL includes all the best practices and therefore can be used to effectively model this type of processes. Traceability between these references and each of the language constructs is provided in every language section.

We have finally selected 7 recent scientific references that discuss the ML lifecycle and its challenges in actual applications \citep{Amershi,Ashmore,Haakman,Kreuzberger,Lwakatare,Nascimento,Shankar}; and 1 paper that blueprints a maturity framework for ML model management \citep{Akkiraju}.

Complementary to the scientific literature, we have chosen 4 industrial publications. First, CRISP-DM~\citep{CRISP} as the de facto standard process for data science projects. Since AI software is, in fact, data-intensive software \citep{Amershi,Khomh,Tamburri}, several practitioners have embarked on their AI projects with an adapted version of CRISP-DM \citep{Amershi,Martinez,Schroer,Tamburri}. We have also selected the Microsoft Team Data Science Process~\citep{TDSP} (henceforth, TDSP), Google Cloud MLOps~\citep{GoogleCloud}, and IBM AI Model Lifecycle Management~\citep{IBM} (henceforth, AIMLM) as three major players in the field. We have additionally included a book that describes the ML lifecycle along with a series of technical considerations and best practices \citep{Burkov} because of its holistic overview.

Each of those proposals has a slightly different clustering, distribution and granularity of activities, but they all share the following high-level structure:
\begin{enumerate}
    \item \emph{Business Understanding}, to analyze the current situation, set the business objectives and criteria, and produce a plan along with an initial assessment of tools and techniques;
    \item \emph{Data Preparation}, to perform the activities that are required to gather and clean data, and prepare data sets and features for creating AI models;
    \item \emph{AI Modeling}, to select AI modeling techniques, optimize hyperparameters and train the AI models, which will be evaluated and ranked according to evaluation and business criteria; and
    \item \emph{Operations}, to finally make the AI models available for consumption and to build a monitoring system and pipelines for continuous improvement.
\end{enumerate}

We would like to note that the literature reviewed focuses on systems with AI components relying on supervised learning techniques, and thus prioritizing activities in which data is intensely processed before being consumed by the AI models. Extensions to the language to cover other types of AI processes such as reinforcement learning are mentioned, but not detailed \citep{Amershi}.
\section{DSL Design}
\label{sec:dsldesign}

In this section, we introduce the proposed DSL for describing ML engineering processes inspired by state-of-the-art industrial and scientific practices (reviewed in Section~\ref{sec:mlenginpractice}).

DSLs generally consist of two main components: an abstract syntax (\emph{i.e.} the language metamodel) and a concrete syntax (\emph{i.e.} the language notation). This section focuses on the abstract syntax, while a visual notation is described as part of the tool support (see Section~\ref{sec:toolkit}).

To facilitate its description, we decompose the abstract syntax in several subsections. The first one (Section~\ref{sec:coreprocessmodelingconcepts}) covers the core elements of any development process, and therefore the elements common to other process modeling languages. Then, in subsequent sections we start extending these core elements with new subclasses and subsystems to model specific aspects of ML engineering such as the new AI roles (Section~\ref{sec:airoles}), AI artifacts and resources (Section~\ref{sec:aiartifactsresources}), and AI-oriented activities (Section~\ref{sec:aiactivities}).

For each package, we will show and explain the key metamodel excerpts. Nonetheless, a complete version of the DSL is available online.


\subsection{Core Process Modeling Concepts}
\label{sec:coreprocessmodelingconcepts}
 
\begin{figure*}[h]
\includegraphics[width=\textwidth]{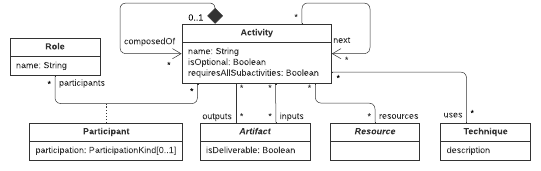}
\caption{Generic elements of an activity.}
\label{fig:Activity}
\end{figure*}

As shown in Figure \ref{fig:Activity}, at its core, our DSL contains the generic concept of activity, their relationship and the main elements they are related to.

An \emph{Activity} constitutes the key element of any process, and it is commonly defined as a piece of work {\textendash}to be performed either manually or automatically{\textendash} that forms one logical step within a process. Activities might be composed of other activities, resulting in a work-breakdown structure (association \emph{composedOf}). Completing an activity may require completing all sub-activities (attribute \emph{requiresAllSubactivities}). Process creators define if an activity is mandatory (attribute \emph{isOptional}). The association \emph{next} provides a construct for a precedence relationship between activities. We do not impose a specific structure in this sense (such as a sequential ordering) but let the authors establish their preferred structure; thus the \emph{*} multiplicity. For complex flows and gateways, the DSL could be accompanied by a model described in any behavioral notation like BPMN.

Activities consume (\emph{input}) and produce (\emph{output}) \emph{Artifacts}. An artifact could be a document that is generated as an output of an activity and is consumed as an input by the following one. Industrial methods are more precise in identifying which artifacts should be generated as output of their activities, and describe their rationale. We will see other examples of artifacts in the next sections. A \emph{Resource} might be helpful to complete an activity. Resources are not consumed nor produced {\textendash}they are supporting components that help the realization of activities. Examples of resources are, for instance, references to public templates to use for generating output documentation and guidelines indicating which are the common techniques recommended being applied in specific activities.

Activities are performed by \emph{Roles}, who are experts in different \emph{Techniques}. Their \emph{participation} could be specified according to the organization's levels of responsibility, \emph{e.g.}, as responsible or accountable (class \emph{Participant}). To perform an activity, it may be required to use techniques.


\subsection{AI Artifacts and Resources}
\label{sec:aiartifactsresources}

In this section, we define elements to include artifacts and resources as auxiliary assets that are managed, used and/or consumed within the activities of an ML-based software development process. The artifacts are created and updated throughout the whole development process, whereas the resources are physical/logical items and knowledge that is available for the company to guide and help the completion of activities.

\begin{figure}[h]
\includegraphics[width=\textwidth]{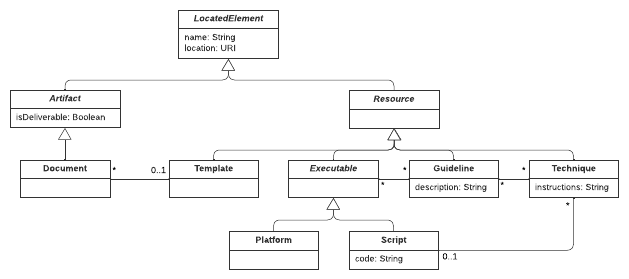}
\caption{An excerpt of the artifacts and resources hierarchy.}
\label{fig:LocatedElements}
\end{figure}

Figure~\ref{fig:LocatedElements} depicts an excerpt of those elements, particularly some resources for supporting the realization of activities such as \emph{Templates} for building \emph{Documents}, \emph{Techniques} to follow during the activities (\emph{e.g.}, algorithms, design patterns, etc.), or \emph{Guidelines} to provide further information and descriptions of procedures and executable \emph{Scripts}.

A key artifact that is particular to AI processes is an \emph{AIModel}: an algorithm that provides end users with automated logical decision-making. \emph{AIModels} are built, trained and evaluated on top of \emph{Datasets} of the \emph{Data} that is available and collected from diverse \emph{DataSources}. Many specific \emph{Techniques} are known to address a very wide range of problems that are commonly solved by \emph{AIModels} {\textendash}from classification, regression and forecasting of discrete data to other unstructured data-based situations such as those in the areas of computer vision, voice recognition or real-time language translation.

The methods reviewed designate specific artifacts 
and briefly expose their intention and contents. Particularly, TDSP includes numerous templates for most of their artifacts. The artifacts found in the literature 
were of very diverse nature; for instance, we see a variety of documents to describe the data that is consumed; software components such as data sets and ML models; and other supporting elements like criteria, guidelines, requirements, hyperparameters, techniques, and a dashboard for telemetry.

Table~\ref{tab:MethodsLocatedElements} lists all artifacts and resources of the DSL and links them to the methods where they are either identified, briefly overviewed or exhaustively defined. We can observe that the industrial methods are very detailed in identifying and describing the expected artifacts and resources to use during the whole process. Meanwhile, the scientific literature is more agnostic and tends to overview necessities in that regard. Our DSL covers all the aspects found in the selected references.

\begin{table*}[h]
\centering
\caption{Mapping of artifacts and resources defined in the DSL that cover referenced methods and studies.}
\label{tab:MethodsLocatedElements}
\begin{tabular}{|l|l|*{13}{c|}}
\hline
    \multicolumn{2}{|c|}{}
    & \multicolumn{8}{|c|}{Scientific literature}
    & \multicolumn{4}{|c|}{Industry}
    & \\
\hline
    \multicolumn{2}{|c|}{}
    & \rotatebox{90}{Akkiraju et al}
    & \rotatebox{90}{Amershi et al}
    & \rotatebox{90}{Ashmore et al}
    & \rotatebox{90}{Haakman et al}
    & \rotatebox{90}{Kreuzberger et al}
    & \rotatebox{90}{Lwakatare et al}
    & \rotatebox{90}{Nascimento et al}
    & \rotatebox{90}{Shankar et al}
    & \rotatebox{90}{CRISP-DM}
    & \rotatebox{90}{Google Cloud MLOps}
    & \rotatebox{90}{IBM AIMLM}
    & \rotatebox{90}{Microsoft TDSP}
    & \rotatebox{90}{Burkov}\\
\hline
    \multirow{4}{5em}{Artifact} &
    Document
    & \(\bullet\) 
    & 
    & \(\bullet\) 
    & \(\bullet\) 
    & \(\bullet\) 
    & 
    & 
    & 
    & \(\bullet\) 
    & \(\bullet\) 
    & 
    & \(\bullet\) 
    & 
    \\
    & Data
    & 
    & 
    & 
    & \(\bullet\) 
    & \(\bullet\) 
    & \(\bullet\) 
    & 
    & \(\bullet\) 
    & \(\bullet\) 
    & \(\bullet\) 
    & \(\bullet\) 
    & \(\bullet\) 
    & \(\bullet\) 
    \\
    & AI Model
    & \(\bullet\) 
    & 
    & \(\bullet\) 
    & 
    & \(\bullet\) 
    & \(\bullet\) 
    & 
    & \(\bullet\) 
    & \(\bullet\) 
    & \(\bullet\) 
    & \(\bullet\) 
    & \(\bullet\) 
    & \(\bullet\) 
    \\
    & AI Model Dataset
    & \(\bullet\) 
    & \(\bullet\) 
    & \(\bullet\) 
    & \(\bullet\) 
    & \(\bullet\) 
    & 
    & 
    & \(\bullet\) 
    & \(\bullet\) 
    & \(\bullet\) 
    & \(\bullet\) 
    & \(\bullet\) 
    & \(\bullet\) 
    \\
\hline
    \multirow{5}{5em}{Resource} &
    Template
    & 
    & 
    & 
    & 
    & 
    & 
    & 
    & 
    & 
    & \(\bullet\) 
    & 
    & \(\bullet\) 
    & 
    \\
    & Data Source
    & 
    & \(\bullet\) 
    & 
    & 
    & \(\bullet\) 
    & 
    & 
    & 
    & \(\bullet\) 
    & 
    & \(\bullet\) 
    & \(\bullet\) 
    & 
    \\
    & Executable
    & \(\bullet\) 
    & \(\bullet\) 
    & \(\bullet\) 
    & 
    & \(\bullet\) 
    & \(\bullet\) 
    & 
    & 
    & \(\bullet\) 
    & \(\bullet\) 
    & \(\bullet\) 
    & \(\bullet\) 
    & 
    \\
    & Guideline
    & 
    & 
    & 
    & 
    & \(\bullet\) 
    & 
    & 
    & 
    & 
    & \(\bullet\) 
    & \(\bullet\) 
    & \(\bullet\) 
    & 
    \\
    & Technique
    & \(\bullet\) 
    & \(\bullet\) 
    & \(\bullet\) 
    & 
    & \(\bullet\) 
    & 
    & 
    & \(\bullet\) 
    & \(\bullet\) 
    & \(\bullet\) 
    & \(\bullet\) 
    & \(\bullet\) 
    & \(\bullet\) 
    \\
\hline
\end{tabular}
\end{table*}


\subsection{AI Roles}
\label{sec:airoles}
\begin{figure*}[h]
\includegraphics[width=\textwidth]{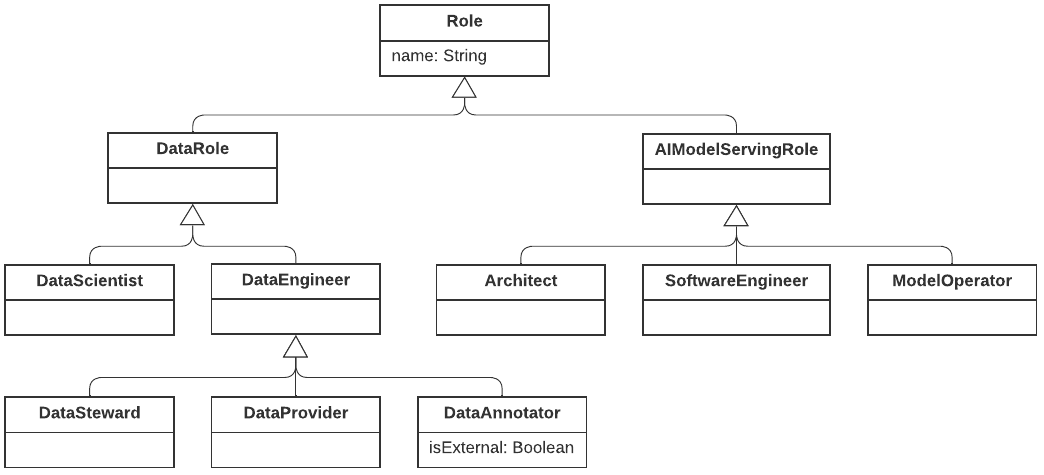}
\caption{The AI-specific role hierarchy.}
\label{fig:RolesHierarchy}
\end{figure*}

A diverse set of specialized skills and functions is required in order to properly and efficiently execute the different activities. In this section, we present the list of roles resulting from the review performed in Section~\ref{sec:mlenginpractice}, and describe their responsibilities and participation within an ML-based software development process.

As in any software development environment, it is important to clarify the  
roles concerned with the objectives and the management of teams and processes, and software architects and engineers. We find new roles related to AI activities, that we present in Figure~\ref{fig:RolesHierarchy}.

There are some roles related to product or project management within an organization that are basically enumerated and have some high-level technical functions assigned. Furthermore, considering the domain an organization is operating in, we identify the roles \emph{DataConsumer}, \emph{BusinessUser} and \emph{BusinessAnalyst}.

In regard to AI activities to be performed largely for acquiring and preparing data, we identify \emph{DataEngineers} and \emph{DataScientists}. Both roles are broadly mentioned in most of the literature. Mainly, data engineers are specialized in collecting, processing, cleaning and harmonizing data; and data scientists analyze, group and filter data, and apply AI algorithms and techniques for creating AI models. We make explicit three subtypes of specialized data engineers with a subset of functions. \emph{DataProviders} make the data available for other stakeholders, by collecting it from various data sources in different locations. On the other hand, \emph{DataStewards} are those responsible for maintaining and implementing the data governance policies of the company, and as such are able to grant access to data assets. A \emph{DataAnnotator} is someone who is tasked with annotating or labelling data. People with this role could be a member of the organization that is running the AI project, or could be externalized as contractors or crowdsourced.

Finally, we have \emph{Architects} and/or \emph{SoftwareEngineers} who help integrate the AI models within the system and release them to production. \emph{ModelOperators} are responsible for monitoring and making sure that an AI model is performing as expected. A mix of the aforementioned roles is found as DevOps Engineer or MLOps Engineer.

When cataloging and defining the roles and functions of the team members for each of the activities, 
some methods zoom in on concrete activities and others keep a generic overview. For instance, AIMLM clearly defines a series of roles and functions for data-related activities, such as Data Consumers, Data Providers, and Data Stewards {\textendash}that could be considered different nuances of the commonly known roles of Data Scientists and Data Engineers. Along with the Data Annotator role from \citet{Akkiraju} and \citet{Burkov}, we see the variety of role definitions according to the particularities of the job to perform only for data engineering profiles.

Overall, there is a myriad of roles and overlapping functions in the literature that we have reviewed. For instance, \citet{Ashmore} describes the functions of an ML Engineer as the same as other authors have designated for Data Scientists, whereas an ML Engineer featured in \citet{Kreuzberger} is a cross-functional mixture of several other profiles, namely: Data Scientists, Data Engineers, Software Engineers, and Model Operators. We consider this as a reflection of the actual practice in the industry, \emph{i.e.}, every company has its own set of job positions, and they are as diverse as each company's organizational structure.

Table~\ref{tab:MethodsRoles} traces the roles to the reviewed methods proposing them, or that include a role with a different name but a similar function or subset of functions. Our DSL includes entities for modeling all the functions referenced in the literature.

\begin{table*}[h]
\centering
\caption{Mapping of roles of the DSL versus methods and studies they cover.}
\label{tab:MethodsRoles}
\begin{tabular}{|l|l|*{13}{c|}}
\hline
    \multicolumn{2}{|c|}{}
    & \multicolumn{8}{|c|}{Scientific literature}
    & \multicolumn{4}{|c|}{Industry}
    & \\
\hline
    \multicolumn{2}{|c|}{}
    & \rotatebox{90}{Akkiraju et al}
    & \rotatebox{90}{Amershi et al}
    & \rotatebox{90}{Ashmore et al}
    & \rotatebox{90}{Haakman et al}
    & \rotatebox{90}{Kreuzberger et al}
    & \rotatebox{90}{Lwakatare et al}
    & \rotatebox{90}{Nascimento et al}
    & \rotatebox{90}{Shankar et al}
    & \rotatebox{90}{CRISP-DM}
    & \rotatebox{90}{Google Cloud MLOps}
    & \rotatebox{90}{IBM AIMLM}
    & \rotatebox{90}{Microsoft TDSP}
    & \rotatebox{90}{Burkov}\\
\hline
    \multirow{3}{7em}{Management Role} &
    Group Manager
    & 
    & 
    & 
    & 
    & 
    & 
    & 
    & 
    & 
    & 
    & 
    & \(\bullet\) 
    & 
    \\
    & Team Lead
    & 
    & 
    & 
    & 
    & 
    & 
    & 
    & 
    & 
    & 
    & 
    & \(\bullet\) 
    & 
    \\
    & Project Lead
    & \(\bullet\) 
    & 
    & 
    & \(\bullet\) 
    & \(\bullet\) 
    & 
    & 
    & 
    & \(\bullet\) 
    & 
    & 
    & \(\bullet\) 
    & 
    \\
\hline
    \multirow{3}{7em}{Domain Role} &
    Data Consumer
    & 
    & 
    & 
    & 
    & 
    & 
    & 
    & 
    & 
    & 
    & \(\bullet\) 
    & 
    & 
    \\
    & Business User
    & 
    & \(\bullet\) 
    & 
    & \(\bullet\) 
    & 
    & 
    & 
    & 
    & 
    & 
    & \(\bullet\) 
    & 
    & 
    \\
    & Business Analyst
    & \(\bullet\) 
    & 
    & 
    & 
    & 
    & 
    & 
    & 
    & \(\bullet\) 
    & 
    & 
    & \(\bullet\) 
    & \(\bullet\) 
    \\
\hline
    \multirow{5}{7em}{Data Role} &
    Data Engineer
    & \(\bullet\) 
    & \(\bullet\) 
    & 
    & 
    & \(\bullet\) 
    & 
    & 
    & 
    & 
    & 
    & \(\bullet\) 
    & \(\bullet\) 
    & \(\bullet\) 
    \\
    & Data Steward
    & 
    & 
    & 
    & 
    & 
    & 
    & 
    & 
    & 
    & 
    & \(\bullet\) 
    & 
    & 
    \\
    & Data Provider
    & 
    & 
    & 
    & \(\bullet\) 
    & 
    & 
    & 
    & 
    & 
    & 
    & \(\bullet\) 
    & 
    & 
    \\
    & Data Annotator
    & \(\bullet\) 
    & \(\bullet\) 
    & 
    & 
    & 
    & 
    & 
    & \(\bullet\) 
    & 
    & 
    & 
    & 
    & \(\bullet\) 
    \\
    & Data Scientist
    & \(\bullet\) 
    & \(\bullet\) 
    & \(\bullet\) 
    & \(\bullet\) 
    & \(\bullet\) 
    & \(\bullet\) 
    & 
    & \(\bullet\) 
    & \(\bullet\) 
    & \(\bullet\) 
    & \(\bullet\) 
    & \(\bullet\) 
    & 
    \\
\hline
    \multirow{3}{7em}{AI Model Serving Role} &
    Architect
    & 
    & 
    & 
    & 
    & \(\bullet\) 
    & 
    & 
    & 
    & 
    & 
    & 
    & \(\bullet\) 
    & 
    \\
    & Software Engineer
    & \(\bullet\) 
    & 
    & 
    & 
    & \(\bullet\) 
    & 
    & 
    & 
    & 
    & 
    & \(\bullet\) 
    & 
    & 
    \\
    & Model Operator
    & \(\bullet\) 
    & 
    & 
    & 
    & \(\bullet\) 
    & 
    & 
    & 
    & 
    & 
    & \(\bullet\) 
    & 
    & \(\bullet\) 
    \\
\hline
\end{tabular}
\end{table*}


\subsection{AI Activities}
\label{sec:aiactivities}
\begin{figure*}[h]
\includegraphics[width=\textwidth]{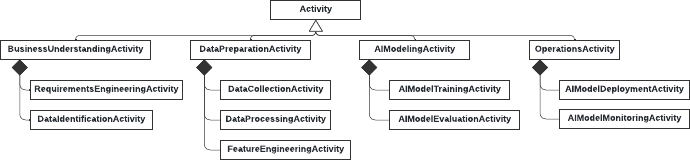}
\caption{High-level view of activities and sub-activities.}
\label{fig:ProcessOverview}
\end{figure*}

Based on the analysis of existing literature, we predefine four main AI-related activities extending the core Activity concept (see Figure~\ref{fig:ProcessOverview}): (1) \emph{BusinessActivity}, (2) \emph{DataPreparationActivity}, (3) \emph{AIModelingActivity}, and (4) \emph{OperationsActivity}.

An AI process may begin with an assessment of the current business situation during \emph{BusinessActivity}, followed by a second group of activities within \emph{DataActivity} to collect and process data in order to create, train and evaluate AI predictive models, which happens during \emph{AIModelingActivity}. Finally, the AI models that match the criteria are deployed and continuously monitored on production during \emph{OperationsActivity}. Each of the high-level activities is composed by a second level group of activities; the execution order of sub-activities within a higher order one might be refined by an adjacent behavioral model.

The granularity of the activities identified for each stage depends on how prescriptive is the corresponding method. For instance, CRISP-DM focuses on data engineering activities and therefore its identification and description of low-level granular activities in this stage, in contrast to other aspects such as operationalization. Indeed, \citet{Schroer} highlights CRISP-DM's lack of guidance towards the deployment phase of any given project.

\citet{Burkov}, CRISP-DM, Google Cloud, TDSP and AIMLM provide recommendations and best practices to perform their activities. TDSP includes workflows that depict the steps to follow, whereas CRISP-DM, AIMLM and others describe the activities in narrative. Google Cloud, TDSP and AIMLM, as vendor-specific methods, also comprehend a series of low level technical tasks and some considerations for their proprietary platforms. Scientific literature, on the other hand, are more generic in their descriptions.

It is important to note that the technical tasks required to complete an activity are extremely dependent on the nature of the problem to address, as well as the specificity of the method. For instance, TDSP presents a series of tasks to perform for cleaning and wrangling text data sets, and includes an excerpt of techniques for each of those tasks. On the other hand, \citet{Amershi} keeps a high level stake on describing their proposal of activities.

We provide a work breakdown structure of activities as a result of a bottom-up analysis of the activities and tasks described in the reviewed literature \textendash{}see Table~\ref{tab:MethodsDimensions} for the covering of each of them. As a result, our DSL contains a super set of activity definitions that comprehends the union of all the referenced methods.

\begin{table*}[h]
\centering
\caption{Mapping of activities of the DSL and the references they are a super set of.}
\label{tab:MethodsDimensions}
\begin{tabular}{|l|l|*{13}{c|}}
\hline
    \multicolumn{2}{|c|}{}
    & \multicolumn{8}{|c|}{Scientific literature}
    & \multicolumn{4}{|c|}{Industry}
    & \\
\hline
    \multicolumn{2}{|c|}{}
    & \rotatebox{90}{Akkiraju et al}
    & \rotatebox{90}{Amershi et al}
    & \rotatebox{90}{Ashmore et al}
    & \rotatebox{90}{Haakman et al}
    & \rotatebox{90}{Kreuzberger et al}
    & \rotatebox{90}{Lwakatare et al}
    & \rotatebox{90}{Nascimento et al}
    & \rotatebox{90}{Shankar et al}
    & \rotatebox{90}{CRISP-DM}
    & \rotatebox{90}{Google Cloud MLOps}
    & \rotatebox{90}{IBM AIMLM}
    & \rotatebox{90}{Microsoft TDSP}
    & \rotatebox{90}{Burkov}\\
\hline
    \multirow{2}{6em}{Business} & 
    Reqs. Engineering
    & 
    & \(\bullet\) 
    & 
    & 
    & \(\bullet\) 
    & \(\bullet\) 
    & \(\bullet\) 
    & 
    & \(\bullet\) 
    & 
    & \(\bullet\) 
    & \(\bullet\) 
    & 
    \\
    & Data Identification
    & \(\bullet\) 
    & 
    & 
    & \(\bullet\) 
    & \(\bullet\) 
    & 
    & 
    & 
    & \(\bullet\) 
    & 
    & \(\bullet\) 
    & \(\bullet\) 
    & 
    \\
\hline
    \multirow{3}{6em}{Data Preparation} &
    Data Collection
    & \(\bullet\) 
    & \(\bullet\) 
    & \(\bullet\) 
    & 
    & \(\bullet\) 
    & \(\bullet\) 
    & \(\bullet\) 
    & 
    & \(\bullet\) 
    & \(\bullet\) 
    & \(\bullet\) 
    & \(\bullet\) 
    & \(\bullet\) 
    \\
    & Data Processing
    & \(\bullet\) 
    & \(\bullet\) 
    & \(\bullet\) 
    & \(\bullet\) 
    & \(\bullet\) 
    & 
    & \(\bullet\) 
    & \(\bullet\) 
    & \(\bullet\) 
    & \(\bullet\) 
    & \(\bullet\) 
    & \(\bullet\) 
    & \(\bullet\) 
    \\
    & Feature Engineering
    & \(\bullet\) 
    & \(\bullet\) 
    & 
    & 
    & \(\bullet\) 
    & 
    & \(\bullet\) 
    & 
    & 
    & \(\bullet\) 
    & \(\bullet\) 
    & \(\bullet\) 
    & \(\bullet\) 
    \\
\hline
    \multirow{2}{6em}{AI Modeling} &
    AI Model Training
    & \(\bullet\) 
    & \(\bullet\) 
    & \(\bullet\) 
    & \(\bullet\) 
    & \(\bullet\) 
    & \(\bullet\) 
    & \(\bullet\) 
    & \(\bullet\) 
    & \(\bullet\) 
    & \(\bullet\) 
    & \(\bullet\) 
    & \(\bullet\) 
    & \(\bullet\) 
    \\
    & AI Model Evaluation
    & \(\bullet\) 
    & \(\bullet\) 
    & \(\bullet\) 
    & \(\bullet\) 
    & \(\bullet\) 
    & \(\bullet\) 
    & \(\bullet\) 
    & \(\bullet\) 
    & \(\bullet\) 
    & \(\bullet\) 
    & \(\bullet\) 
    & \(\bullet\) 
    & \(\bullet\) 
    \\
\hline
    \multirow{2}{6em}{Operations} &
    AI Model Deployment
    & \(\bullet\) 
    & \(\bullet\) 
    & \(\bullet\) 
    & 
    & \(\bullet\) 
    & \(\bullet\) 
    & \(\bullet\) 
    & \(\bullet\) 
    & 
    & \(\bullet\) 
    & \(\bullet\) 
    & \(\bullet\) 
    & \(\bullet\) 
    \\
    & AI Model Monitoring
    & \(\bullet\) 
    & \(\bullet\) 
    & \(\bullet\) 
    & \(\bullet\) 
    & \(\bullet\) 
    & \(\bullet\) 
    & \(\bullet\) 
    & \(\bullet\) 
    & 
    & \(\bullet\) 
    & \(\bullet\) 
    & 
    & \(\bullet\) 
    \\
\hline
\end{tabular}
\end{table*}

In the following, we present the different activities and related elements of the DSL. Each organization will then be able to pick from the list those that are important for their specific process.


\subsubsection{Business Understanding Activity}
Software is the means towards achieving business results that, aligned to the strategy and culture of an organization, describe its success in their domain. Setting a realistic goal and describing a comprehensive context is a key action for determining the expected impact and value provided by any internal initiative.

\begin{figure*}[h]
\centering
\includegraphics[width=0.8\textwidth]{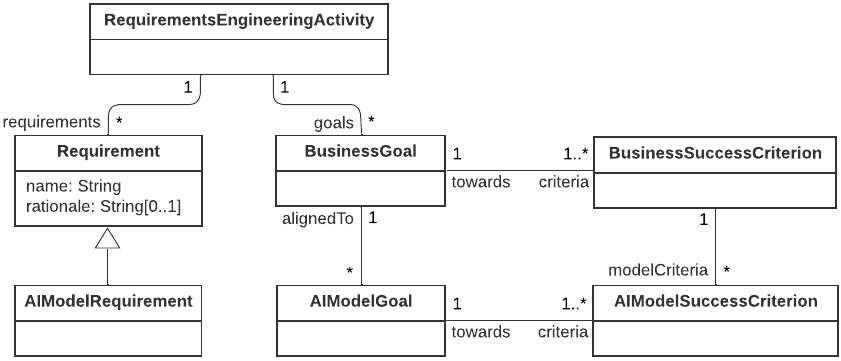}
\caption{An excerpt of components of the requirements engineering activity.}
\label{fig:RequirementsEngineeringActivity}
\end{figure*}

The \emph{RequirementsEngineeringActivity} (see Figure~\ref{fig:RequirementsEngineeringActivity}) is intended for defining and agreeing on different aspects of an initiative. A set of \emph{BusinessGoals} must be defined to provide a sense of purpose and be able to qualitatively describe the success of an AI initiative. Those business goals would be evaluated by a set of quantitative \emph{BusinessSuccessCriteria} that should illustrate how a successful software system should look like. An example of a business goal could be \emph{"reduce user churn"}. A success criterion to evaluate it would be \emph{"reduce the number of users with 4-week inactivity from current 60\% to 40\%"}. The attributes \emph{baseline} and \emph{target} from the superclass \emph{SuccessCriterion} express the different elements of a \emph{BusinessSuccessCriterion}. The attribute \emph{dataType} states which is the unit for both \emph{baseline} and \emph{target}.

Framing a product or business goal in terms of an AI problem facilitates aligning expectations from stakeholders and agreeing on the suitability of an AI initiative. Correlated to the business goals, then, are the \emph{AIModelGoals}, which should be {\textendash}under the responsibility of the AI software component{\textendash} drivers towards the success of the global initiative. As a result of this activity also a set of \emph{AIModelSuccessCriteria} must be defined in order to quantitatively evaluate the performance and contribution of the AI models towards achieving their goals.

A set of \emph{Requirements} (and assumptions and constraints) should be identified to evaluate the resulting system by other qualitative and quantitative factors. Some of those requirements could be also factors that are intrinsic to the intended models, and are classified as \emph{AIModelRequirements}. Those factors could be the tolerated level of degradation of an AI model in production, and a set of ethical requirements, such as fairness.
\begin{figure*}[h]
\centering
\includegraphics[width=\textwidth]{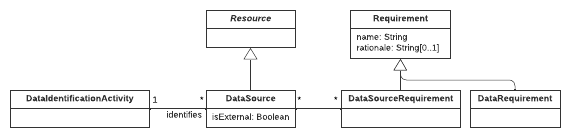}
\caption{An excerpt of components of the data identification activity.}
\label{fig:DataIdentificationActivity}
\end{figure*}

A fundamental result of the \emph{BusinessActivity} is the identification of data sources from which to gather raw data that would guide the rest of the development. This is done during the \emph{DataIdentificationActivity} (see Figure~\ref{fig:DataIdentificationActivity}), in which a series of \emph{DataSources} are discovered and a subset of them are finally selected as the sources for the project. Data reside in specific locations which the development team should connect with to collect it. Data sources could be either internal {\textendash}\emph{e.g.}, the organization's proprietary databases{\textendash}, or external, such as public data sources (attribute \emph{isExternal}). We should consider that accessing and extracting data from those data sources might be constrained by a series of \emph{DataSourceRequirements}, \emph{e.g.}, legal aspects and privacy regulations. A set of \emph{DataRequirements} is also defined to drive the activities for processing and transforming data.


\subsubsection{Data Preparation Activity}
\begin{figure*}[h]
\centering
\includegraphics[width=0.85\textwidth]{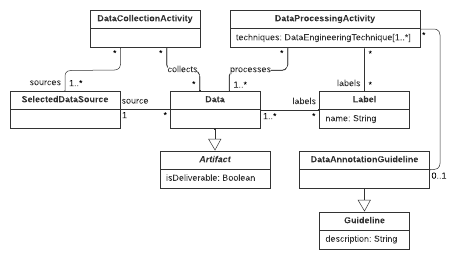}
\caption{An excerpt of components of the data collection and processing activities.}
\label{fig:DataCollectionAndProcessingActivities}
\end{figure*}

The \emph{DataCollectionActivity} (see Figure~\ref{fig:DataCollectionAndProcessingActivities}) is the acquisition of \emph{Data} from the \emph{SelectedDataSources}. The participants may require to apply diverse strategies and technologies to move data from internal or external data sources (attribute \emph{isExternal}) into a destination (attribute \emph{location}) for further processing.

Data may present deficiencies that might result in bad predictions if used for training an AI model. In the \emph{DataProcessingActivity} (see Figure~\ref{fig:DataCollectionAndProcessingActivities}) the data is cleaned and transformed via some data cleaning and transformation \emph{techniques}.

\begin{figure*}[h]
\centering
\includegraphics[width=0.8\textwidth]{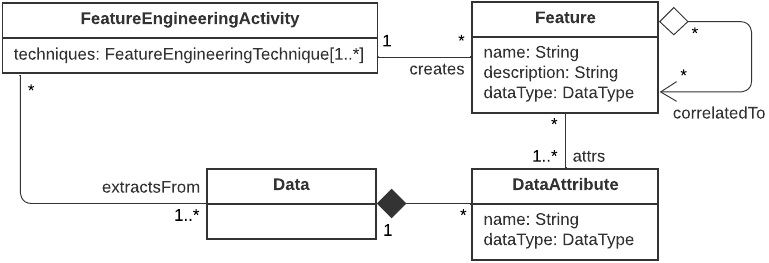}
\caption{An excerpt of components of the feature engineering activity.}
\label{fig:FeatureEngineeringActivity}
\end{figure*}

The \emph{FeatureEngineeringActivity} (see Figure~\ref{fig:FeatureEngineeringActivity}) comprehends the tasks and statistical \emph{techniques} used to select and transform \emph{DataAttributes} of \emph{Data} into features that can be used by an AI model and enhance its prediction accuracy. Feature engineering is an iterative activity where the data scientist creates, extracts, transforms and removes features, looking for the most relevant ones for the problem to address and the most suitable ones for the AI algorithms that will process them. Nonetheless, during this activity, correlations between features' values are identified and expressed (\emph{correlatedTo}).

As a result of this activity, a set of features are included in the data instances. Now the team has a dataset ready for building an AI model, which is split into three different ones with specific intention: a \emph{TrainingDataset}, an evaluation or \emph{ValidationDataset} and a \emph{TestDataset}.


\subsubsection{AI Modeling Activity}
\begin{figure*}[h]
\centering
\includegraphics[width=\textwidth]{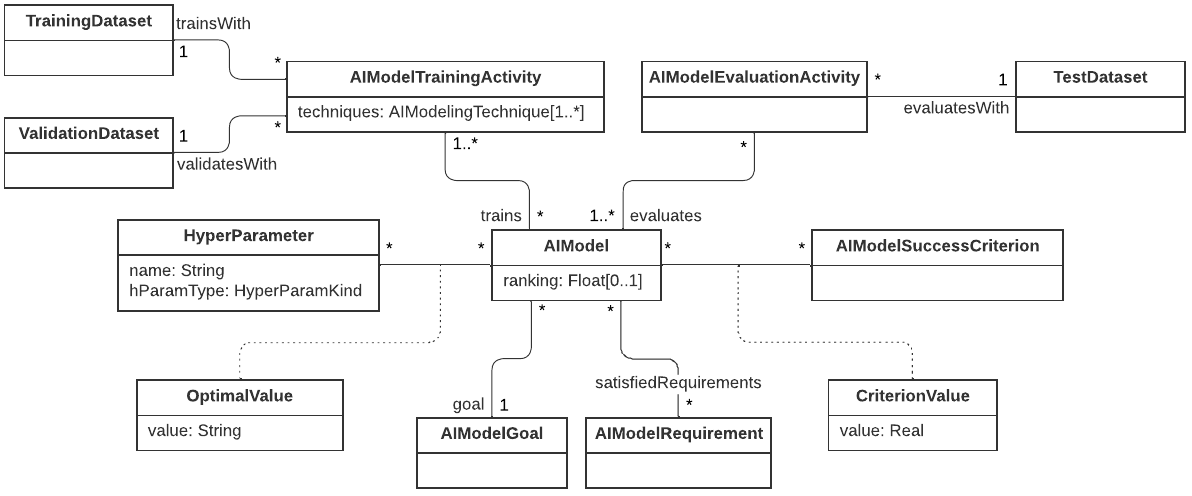}
\caption{An excerpt of activities and components of the AI modeling activity.}
\label{fig:AIModelingActivity}
\end{figure*}
The \emph{AIModelTrainingActivity} (see Figure~\ref{fig:AIModelingActivity}) is the core activity for creating, training and validating new AI models from the collected and prepared data.

An \emph{AIModel} is trained by an AI algorithm (one of the \emph{techniques} informed in the activity) using the observations held in the \emph{TrainingDataset}. Once an AI model is initially trained, a data scientist proceeds to tune its \emph{Hyperparameters} looking for the \emph{OptimalValues} that yield its best performance. The \emph{ValidationDataset} is applied to different AI models or the same AI model configured with different hyperparameter values. Finally, the hyperparameter values that maximize an AI model performance are fixed for production. The \emph{AIModelPerformanceCriteria} will drive the AI model training and will be used to pursue an AI model or discard it; in other words, they dictate when it is not worthwhile to keep improving an AI model.

In the \emph{AIModelEvaluationActivity} (see Figure~\ref{fig:AIModelingActivity}), a data scientist checks if an AI model fits to the \emph{AIModelSuccessCriteria}, along with its adequacy to the \emph{AIModelRequirements}. A \emph{TestDataset} is used to assess the performance of a validated AI model as per those criteria. Data scientists then set a \emph{ranking} for each AI model and evaluate which are the high predictive, the best performers, and most suitable to the AI model requirements to be considered for production.


\subsubsection{Operations Activity}
\begin{figure*}[h]
\centering
\includegraphics[width=\textwidth]{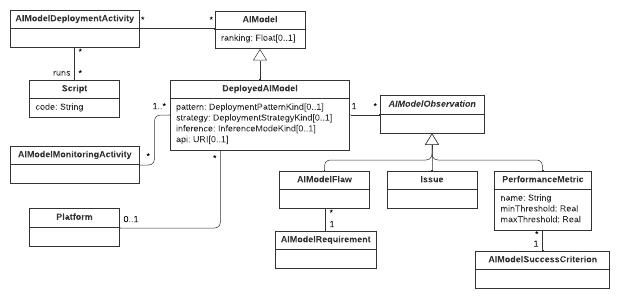}
\caption{An excerpt of activities and components of the operations activity.}
\label{fig:OperationsActivity}
\end{figure*}
The \emph{OperationsActivity} activity describes how the AI model resulting from following the process is deployed 
to be exploited by end users or other systems. Its activities are depicted in Figure~\ref{fig:OperationsActivity}.

In the \emph{AIModelDeploymentActivity}, an AI model is deployed to a production \emph{Platform}. It may be useful to run \emph{Scripts} specific to a technology that semi-automatize the installation and setting up of the configuration properties. An AI model is deployed with a \emph{pattern}. The values of its \emph{DeploymentPatternKind} enumeration type are: (1) statically {\textendash}as part of a software package{\textendash}, (2) dynamically on the user's device, (3) dynamically on a server, and (4) via streaming. Additionally, an AI model is deployed following a \emph{strategy} that helps to decide how and when to release the new predictions to consumption. The enumeration \emph{DeploymentStrategyKind} has the following values: (1) single deployment, (2) silent deployment, (3) canary deployment {\textendash}as an incremental rollout tactic{\textendash} and (4) multi-armed bandit {\textendash}some versions of the same AI model are deployed to different user segments to compare their performance and accuracy in the production environment{\textendash}.

An AI model is set up to make its \emph{inferences} in one of two modes. When serving in batch mode {\textendash}recommended when computation optimization is a priority{\textendash}, the AI model periodically makes predictions offline and stores the results in a repository. On the other hand, if the AI software must prioritize providing near real-time predictions, an AI model should be configured to serve in on-demand mode, so that it makes and delivers predictions whenever requested to. Those values are included in the \emph{InferenceModeKind} enumeration type.

Once an AI model is in production, it is vital to observe and monitor its performance according to the initial \emph{AIModelRequirements} and \emph{AIModelSuccessCriteria}. Operators need to make sure that the AI model is served correctly and performs according to established thresholds. To do that, the system will be continuously collecting interaction data, logs, and payloads.

We consider important to identify the \emph{AIModelFlaws} that should be catched, and which \emph{AIModelRequirement} is a given flaw related to. It is also fundamental to define \emph{PerformanceMetrics} that set the \emph{minThreshold} and \emph{maxThreshold} values that put in place the correct performance of the AI model.
\section{Toolkit}
\label{sec:toolkit}

We have implemented a modeling editor for the DSL presented in Section~\ref{sec:dsldesign} to facilitate the graphical specification of development processes for smart systems. Moreover, additional plug-ins for the editor enable: (1) the export of the modeled process as a BPMN compliant representation for process execution using popular workflow platforms; and (2) an HTML report generator to consolidate, centralize and communicate the information required to understand the company's AI engineering processes and their application. The source code and the installation package of the toolkit are publicly available online\footnote{\url{http://hdl.handle.net/20.500.12004/1/A/MFML/001}}. The next subsections introduce the toolkit components, whereas in Section~\ref{sec:application} we explore an example process, modeled with the editor, and the outputs generated by the plug-ins.


\subsection{Modeling Editor}
\label{sec:modelingeditor}
 
Our modeling editor has been built with Eclipse Sirius\footnote{\url{https://www.eclipse.org/sirius}}, an open-source project hosted at the Eclipse Foundation. Sirius provides a platform to automatically generate modeling tools given a language metamodel and a visual notation. The generated tools can then be used by end-users to model their own processes following the provided visual notation.

\begin{figure*}[t]
\includegraphics[width=\textwidth]{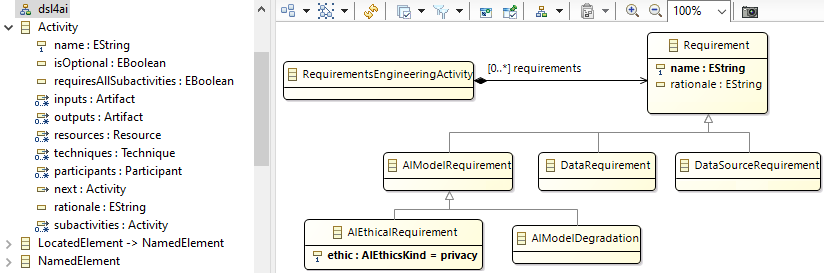}
\caption{A zoom in of the DSL implemented in Sirius.}
\label{fig:SiriusMetamodel}
\end{figure*}

Therefore, we defined all the elements of our DSL abstract syntax and defined a visual notation element for each of them. As in Sirius all metamodels need to have a root node, we have created a new \emph{Method} class for that purpose in the Sirius package containing our DSL, named \emph{dsl4ai}. In Figure \ref{fig:SiriusMetamodel} we show the Sirius modeler editor and an excerpt of our DSL in EMF notation. On the left, there is the list of entities of the \emph{dsl4ai} domain; the entity \emph{Activity} displays its contents, \emph{i.e.}, its attributes and associations with other entities.

\begin{figure*}[h]
\centering
\includegraphics[width=0.6\textwidth]{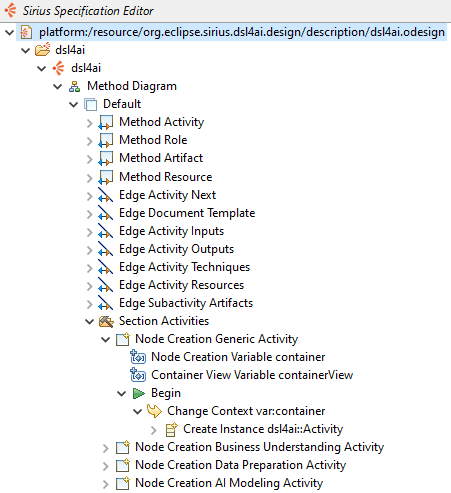}
\caption{The view specification of the \emph{Method Diagram}.}
\label{fig:SiriusViewSpecification}
\end{figure*}

Our modeling editor provides three different views: two for the \emph{Method} and one for any \emph{Activity}. The \emph{Method Diagram} view provides a visual overview of the modelled method, which depicts activities and their sequence relationships, their inputs and outputs, and the roles of the organization. There is a second view called \emph{Method Resources Diagram} that shows all the resources that are used within each high-level activity of the method, namely: guidelines, techniques, scripts, and platforms. Last, the \emph{Activity Diagram} view displays the activities that are contained in a given activity, their sequence relationships, and their referenced artifacts and resources. In Figure~\ref{fig:SiriusViewSpecification} we see an excerpt of the specification of the \emph{Method Diagram}, with elements to visually represent activities, roles, artifacts and resources, and the relationships between all of them. The figure also includes a menu entry point for creating a new activity.

\begin{figure}[h]
\centering
\includegraphics[width=0.4\columnwidth]{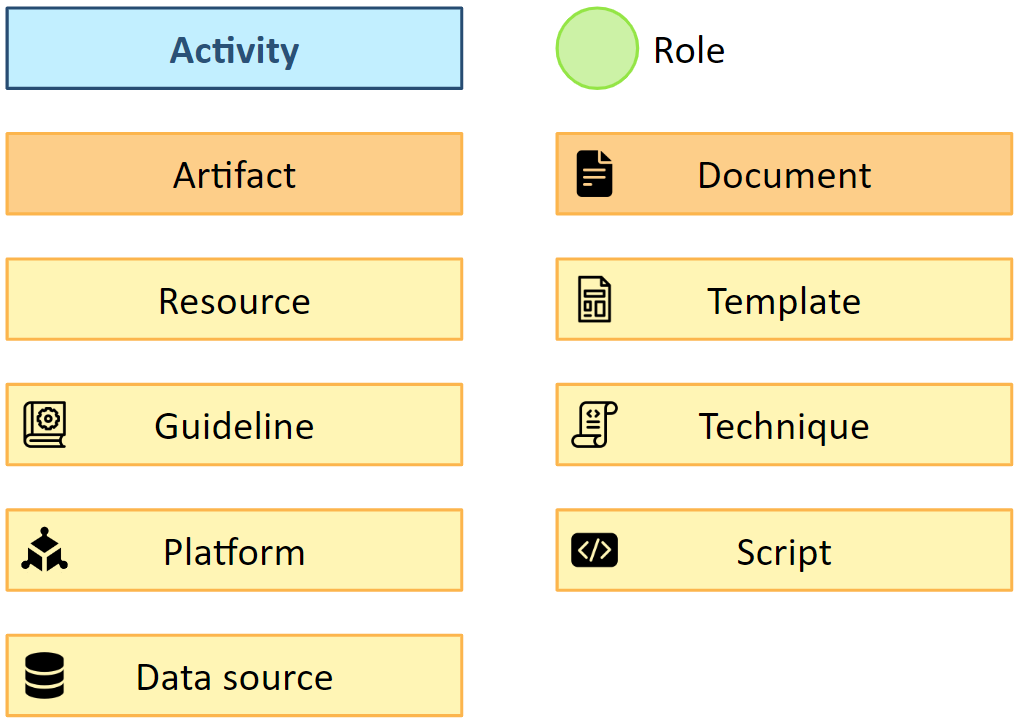}
\caption{Visual notation of the DSL syntax in the modeling editor.}
\label{fig:VisualNotation}
\end{figure}

As we can see in Figure~\ref{fig:VisualNotation}, activities are represented as blue rectangles. Roles are illustrated as green circles. Artifacts are displayed as orange rectangles and resources as yellow ones. Resources embed an icon picturing the specific subclass of the element. The associations establishing a sequence of execution between two activities are represented as a filled, closed output arrow from the source activity to the next one. Associations from an activity to its inputs and outputs are illustrated as arrows so that: an incoming arrow to an activity depicts an input relationship with an artifact, and an outgoing arrow from an activity relates to an output association with an artifact. Links from and to resources are rendered as orange dotted lines.


\subsection{BPMN Conversion and Simulation}
\label{sec:bpmnconverter}
The Business Process Model and Notation (BPMN) is an OMG standard adopted by the industry to model and execute their business processes \citep{Green,zurMuehlen}. As such, being able to convert our DSL to a BPMN process (even if losing details during the transformation) could be beneficial as we could reuse other BPMN tools for particular tasks, such as process simulation and enactment.

To achieve that, we have developed an Eclipse plug-in to export a process built with our modeling editor to a BPMN 2.0 file, so that any process modeler would be able to import, simulate, adapt and execute an AI process in their preferred BPM suite. The Eclipse plug-in has been built as an Acceleo project. Acceleo is an implementation of the MOF Model to Text Transformation Language specification~\cite{MOFM2T} defined by the OMG, and it is composed of two main types of structures (templates and queries) that take a model as an input and generates output code. The main language used for retrieving the model assets is a subset of OCL.

As said above, this transformation implies a simplification of the process modeled using with our DSL, so a direct transformation from our DSL to process simulation tools can be envisioned. Going through BPMN offers right now a good trade-off between the complexity of the transformation and its benefits. 

The following listing shows a snippet of the Acceleo transformation implementing the conversion:

\begin{lstlisting}[caption=Template for generating common BPMN components of an \emph{Activity}.]
[template public generateActivityElements(activity: Activity)]
	[generateActivityDocumentation()/]
	[generateActivityParticipants()/]
	[generateSpecificActivityProperties()/]
	[for (element : NamedElement | activity.inputs.oclAsType(NamedElement)->union(activity.techniques.oclAsType(NamedElement))->union(activity.resources.oclAsType(NamedElement)))]
		[generateDataInputAssociation(activity, element.name, false)/]
	[/for]
	[for (artifact : Artifact | activity.outputs)]
		[generateDataOutputAssociation(activity, artifact.name, false)/]
	[/for]
[/template]
\end{lstlisting}

\subsection{HTML Documentation Generator}
\label{sec:htmlgenerator}
Similarly to the BPMN exporter previously described, we have also included in the toolkit an HTML generator to create an HTML file that reports on all the information contained in a process model in HTML format. The main objective of this feature is for process modelers to publish this information in an online repository of their organization and make this knowledge accessible to the company employees. Anyone at the organization could then browse the definition of the process and navigate throughout all its assets, namely: roles, resources, artifacts, activities (and sub-activities), and other elements presented in the DSL. This Eclipse plug-in is also based on an Acceleo project.

The information is presented as follows. First, a generic element depicting the process and its components: roles, resources (organized by resource subclass), artifacts, high-level activities and all sub-activities defined in the process. Each of the resources and artifacts expose their descriptions and paths (as external links). Activities are links to the respective section in the HTML file with their description.

An \emph{Activity} section displays its rationale, the list of input and output resources and artifacts, and the list of participants. Resources, artifacts, and participants include a link to each corresponding definition {\textendash}included in the process section. This section also includes links to its parent, previous, and next activities; and a link to all its sub-activities.

\section{Application: A TDSP Implementation}
\label{sec:application}

We present in this section a practical example of the usage of the toolkit. We have completely modeled the Microsoft Team Data Science Process (TDSP) {\textendash}including all its activities and roles, artifacts and resources referenced in Microsoft's knowledge base, using the modeling editor{\textendash}, and have run the plug-ins for BPMN export and HTML documentation. The whole example is included as a resource of the toolkit and therefore distributed and available for its use.


\begin{figure*}[h]
\includegraphics[width=\textwidth]{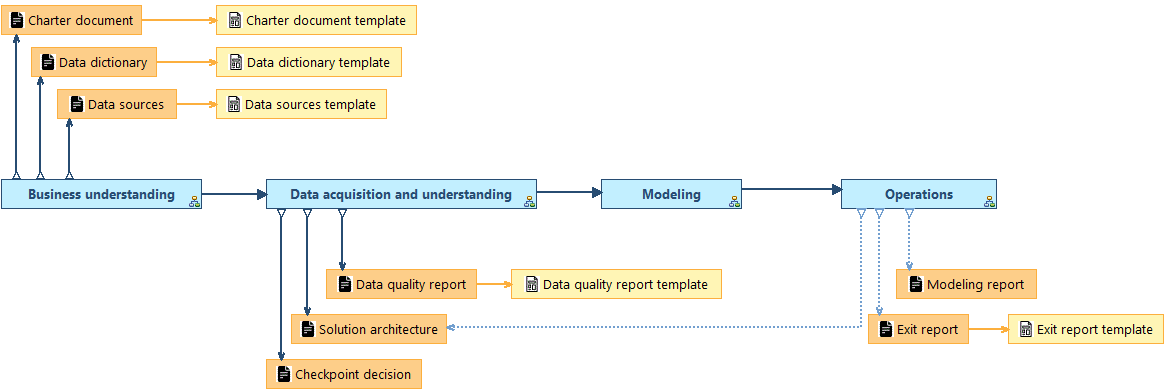}
\caption{The TDSP high-level process designed with the modeling editor.}
\label{fig:SiriusTDSPMethod}
\end{figure*}

The high-level process of TDSP is represented in Figure~\ref{fig:SiriusTDSPMethod}. The TDSP activities \emph{Business understanding}, \emph{Data acquisition and understanding}, \emph{Modeling} and \emph{Deployment} are the instances of \emph{BusinessUnderstandingActivity}, \emph{DataPreparationActivity}, \emph{AIModelingActivity} and \emph{OperationsActivity} of the DSL, respectively. As output artifacts of \emph{Business understanding} we see a \emph{Charter document}, a \emph{Data dictionary}, and a document listing the \emph{Data sources}. All of these documents have an associated template. On the other hand, as a result of the sub-activities of \emph{Deployment}, the method expects a \emph{Modeling report}, a \emph{Solution architecture} (which was the immediate output artifact of \emph{Data acquisition and understanding}), and an \emph{Exit report}.

\begin{figure*}[h]
\includegraphics[width=\textwidth]{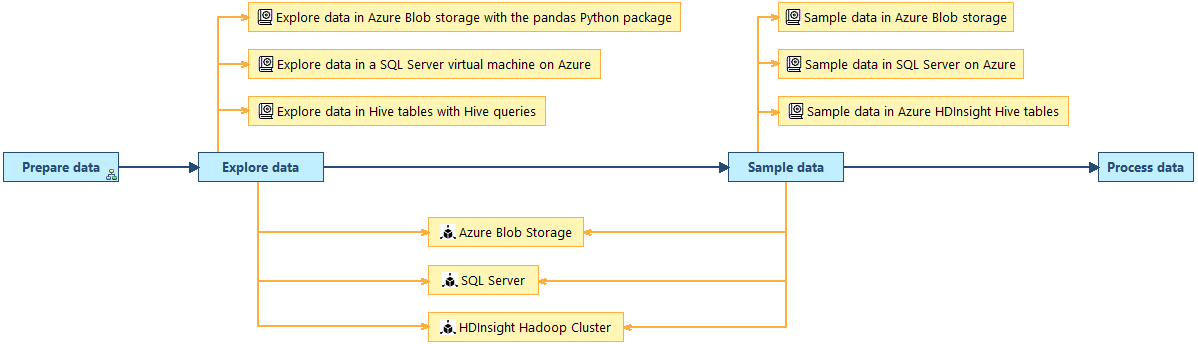}
\caption{The \emph{Explore and visualize data} activity designed with the modeling editor.}
\label{fig:SiriusProcessDataActivity}
\end{figure*}

Figure~\ref{fig:SiriusProcessDataActivity} represents the TDSP activity \emph{Explore and visualize data} which is an instance of the \emph{DataProcessingActivity} of the DSL. We can see that this activity is actually broken down into four sub-activities which have a narrowed task to be completed: \emph{Prepare data}, \emph{Explore data}, \emph{Sample data}, and finally \emph{Process data}. Several guidelines could be used to perform these tasks (\emph{e.g.}, \emph{Explore data in a SQL Server virtual machine on Azure}) and there are some platforms available where to process the data; namely: \emph{Azure Blob Storage}, \emph{SQL Server}, and/or \emph{HDInsight Hadoop Cluster}.

The DSL provides flexibility for adding elements that are specific to a given method, including sub-tasks with a very concrete scope. In the example, the sub-activities of \emph{Explore and visualize data} from TDSP do not have a correspondence with any AI activity described in our DSL. Therefore, they are based on the generic \emph{Activity} entity.
\begin{figure}
\centering
\includegraphics[width=0.5\columnwidth]{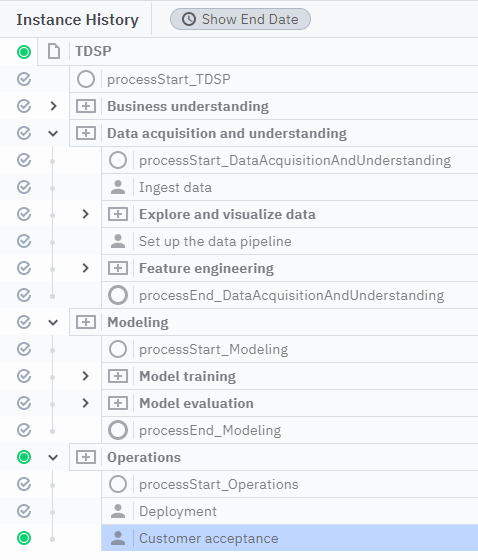}
\caption{A running instance of the TDSP process in Camunda.}
\label{fig:CamundaRunningProcess}
\end{figure}

Moreover, we have generated a BPMN compliant file from the modeled TDSP method and afterward imported it into the Camunda platform\footnote{\url{https://camunda.com}} for modeling and operating business processes. We have run a number of instances of the process to validate the flow is executed as expected. Figure~\ref{fig:CamundaRunningProcess} illustrates the execution history of a TDSP running process instance in Camunda, based on the model imported from our toolkit. It shows the current pending activity is \emph{Customer acceptance}, within the activity \emph{Operations}. All the previous activities and sub-activities have been successfully completed.

\begin{figure*}[h]
\includegraphics[width=\textwidth]{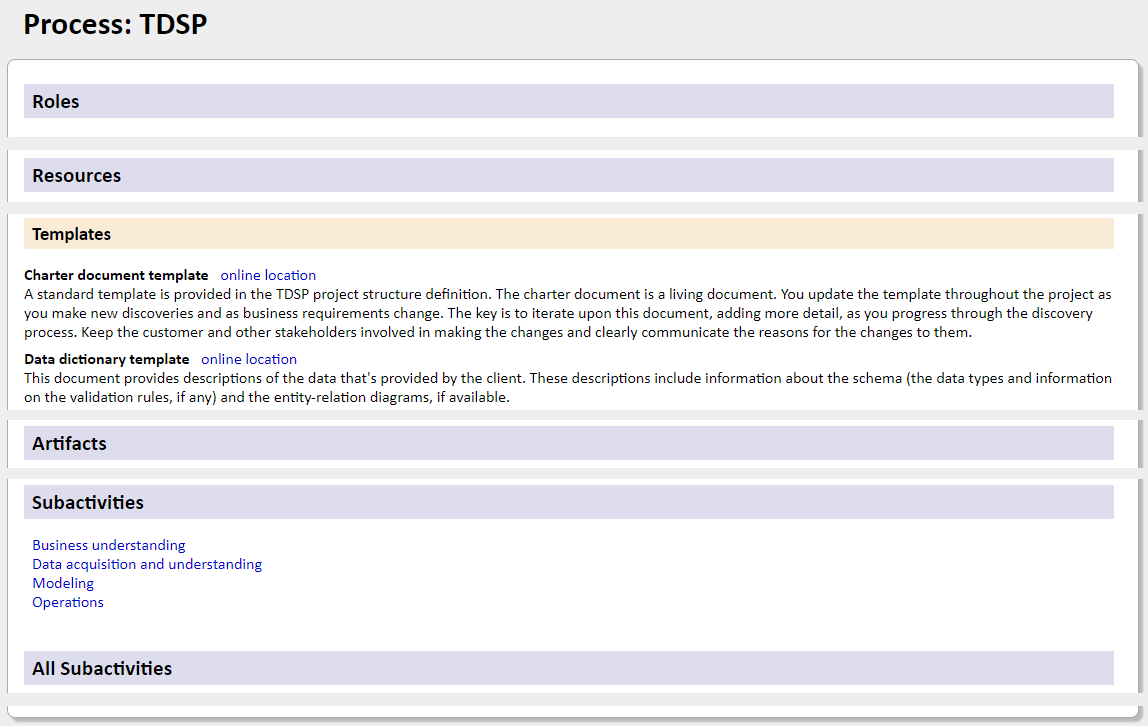}
\caption{An excerpt of the HTML documentation generated from the TDSP method.}
\label{fig:HTMLTDSP}
\end{figure*}

We have also run the HTML documentation generator for that modeled TDSP so that a user would be able to browse online all the activities included in the method, and consult the details modeled for any artifact or resource required for their proper execution. Figure~\ref{fig:HTMLTDSP} shows part of the HTML documentation generated by the feature for the TDSP method. It lists, among other information, the templates that are defined in that method. The high-level process consists of four activities: \emph{Business understanding}, \emph{Data acquisition and understanding}, \emph{Modeling} and \emph{Operations}. Clicking on any of those links will redirect the user to the corresponding activity documentation section.

\section{Related Work}
\label{sec:relatedwork}

Process modeling is a key activity for the proper discovery, documentation, simulation, and enactment of processes, which enable process innovation. \citet{JestonNelis} and \citet{Indulska} enumerated the many benefits of process modeling and process models. First, process modeling facilitates a direct communication between management, the IT team and other staff for the sake of collaboratively supporting the business strategy. Moreover, process models provide team members with an up-to-date guideline on how activities in the organization should be performed according to standard practices and regulations. Simulation and analysis of processes execution bring out bottlenecks and other aspects to consider fixing or improving such processes.

Domain-specific languages are a very useful way to support and boost the essential characteristic of process models {\textendash}to be a mean to profess a shared understanding of the activities to perform, and their related elements. DSLs lay out a set of foundational syntactic and semantic elements that conform to a common language for the business. One that is independent of any technology and is intended to be used across the organization. A visual syntax increments a DSL's communicative expressiveness.

There are dozens of process modeling languages, \emph{e.g.}, BPMN~\citep{BPMN} \& SPEM~\citep{SPEM} and their extensions, UML profiles, and formal languages \citep{Borgonon}. More specifically, SPEM is an OMG standard for describing software development processes, which makes it a clear inspiration for our work. Nevertheless, it purposely does not include any distinct feature for particular domains or disciplines {\textendash}like Artificial Intelligence{\textendash}, asking to extend the core language to provide software process modeling languages for specific types of systems. Even more, to the best of our knowledge, none of the available process modeling languages includes AI specific extensions. Our DSL can be regarded as one of such extensions for ML-based systems. 

Indeed, there is a variety of other DSLs which have been proposed to better specify specific tasks in more traditional development processes. Accordant \citep{Castellanos} is a framework based on a DSL and DevOps practices to help design the software architecture of big data analytics solutions. When it comes to management, integrating existing tools for different process management areas it could be useful to lean on DSLs \citep{Bezivin} by, \emph{e.g.}, keeping traceability between models \citep{Drivalos} or providing process models transformations \citep{Bertero}.

DSLs in the AI field are used to support some AI activities, but not for assisting in the design nor execution of processes involving them. OptiML \citep{OptiML} and ScalOps \citep{ScalOps} optimize the execution of algorithms for data analytics, and Pig Latin \citep{PigLatin} is a language for data querying and processing. Arbiter \citep{Arbiter} expresses ethical requirements for an AI model. ML-Schema \citep{MLSchema} is an ontology for interchanging information on ML experiments which includes classes for datasets, hyperparameter optimization and model evaluation. DeepDSL \citep{DeepDSL} is focused on the creation of deep learning networks, and DEFine \citep{DEFine} allows data scientists to specify, optimize and evaluate deep learning models. Finally, ThingML2 \citep{ThingML2} specializes in the design of IoT components.

As previously mentioned, none of these DSLs cover the dimensions for AI engineering processes discussed in Section~\ref{sec:dsldesign}, considering the particularities of AI activities, roles and artifacts, and their relationship.

\section{Conclusions and Further Work}
\label{sec:conclusions}

In this article, we have proposed a modeling framework to comprehensively specify development processes targeting the creation of smart software systems in an organization.

The framework consists of a DSL that provides a set of mechanisms to express predefined relevant generic aspects of development processes, along with those ML-specific elements for which we define their syntax and semantics. Those assets could then be reused, combined and assembled to design a process model. The framework includes tools to transform a specified process to BPMN and HTML. Since machines can interpret and execute the process syntax in BPMN, any process built with it could be instantiated and conducted within a compatible process execution tool, ensuring standardization and consistency in the dispatching of tasks.

As further work, we plan to extend our DSL to cover other types of AI-based development processes and not just ML ones, such as systems embedding reinforcement learning components. We also envision incorporating process snippets and templates that would help companies to create their own process by composing other method chunks. Finally, we also plan to extend our tool set by implementing additional services that could leverage even more of the modeled processes. For instance, we plan to add code-generation capabilities that could help to link running processes with end-to-end AI pipelines and MLOps systems.

\section{Acknowledgments}
This work has been partially funded by the Spanish government (PID2020-114615RB-I00/AEI/10.13039/501100011033, project LOCOSS), the AIDOaRt project, which has received funding from the ECSEL Joint Undertaking (JU) under grant agreement 101007350, and the Luxembourg National Research Fund (FNR) PEARL program, grant agreement 16544475.

\printcredits

\bibliographystyle{model1-num-names}

\bibliography{references}

\bio{}
\endbio

\bio{}
\endbio

\end{document}